# Reconfigurable Reservoir Computing in a Magnetic Metamaterial

**Vidamour, I. T.[1,2*]; Swindells, C.[1]; Venkat, G.[1]; Manneschi, L.[2]; Fry, P. W.[3]; Welbourne, A.[1]; Rowan-Robinson, R. M.[1]; Backes, D[4]; Maccherozzi, F[4]; Dhesi, S. S.[4]; Vasilaki, E.[2]; Allwood, D. A.[1]; and Hayward, T. J.[1]**

1- Department of Materials Science and Engineering, University of Sheffield, Sheffield, UK
2- Department of Computer Science, University of Sheffield, Sheffield, UK
3- Nanoscience and Technology Centre, University of Sheffield, Sheffield, UK
4- Diamond Light Source, Harwell Science and Innovation Campus, Didcot, UK
*- Corresponding author, contactable at i.vidamour@sheffield.ac.uk

## Abstract

*In-materia reservoir computing (RC) leverages the intrinsic physical responses of functional materials to perform complex computational tasks. Magnetic metamaterials are exciting candidates for RC due to their huge state space, nonlinear emergent dynamics, and non-volatile memory. However, to be suitable for a broad range of tasks, the material system is required to exhibit a broad range of properties, and isolating these behaviours experimentally can often prove difficult. By using an electrically accessible device consisting of an array of interconnected magnetic nanorings- a system shown to exhibit complex emergent dynamics- here we show how reconfiguring the reservoir architecture allows exploitation of different aspects the system's dynamical behaviours. This is evidenced through state-of-the-art performance in diverse benchmark tasks with very different computational requirements, highlighting the additional computational configurability that can be obtained by altering the input/output architecture around the material system.*

## Introduction

In-materia computation, where the responses of material systems are exploited to perform computational operations, offers a potential alternative to conventional CMOS computing. Here, like in biological neurons, data processing operations are performed intrinsically via the physics governing the system's response to inputs. This offers potential improvements in both latency and power efficiency, as dynamical complexity and memory are inherent properties of the substrate. This removes the need to shuttle data between discrete memory and computational units, which can cost up to 100 times the energy of the computation itself when discrete memory units are located off-chip[1].

Reservoir Computing (RC)[2,3] is a bio-inspired computational paradigm which is especially harmonious with *in-materia* computation. In RC, a time-dependent 'reservoir' layer (typically a recurrent neural network, RNN) provides complex nonlinear representations of input data, and a time-invariant readout layer provides a weighted output of the evolving state of the reservoir. Only the readout layer is trained, alleviating the training difficulties associated with standard RNNs since temporal dependencies of the reservoir layer are decoupled from the simple linear output[4].

As the response of the RNN is mathematically analogous to that of a dynamic system, it can be substituted with a real-world dynamic system with appropriate properties, namely nonlinearity between input and output, and a dependence on previous state that asymptotically diminishes over time, termed a 'fading memory'. This has led to a plethora of proposed implementations, with platforms including optoelectronic[5–7], molecular[8], mechanical[9–11], biological[12,13], memristive[14–16], and magnetic[17–23] systems.

Nanomagnetic platforms are of particular interest for RC due to their inherent hysteretic behaviours and nonlinearity of system dynamics, satisfying the two broad criteria necessary for RC. Many magnetic systems have been proposed as reservoirs and come with their own strengths and weaknesses. Spin-torque nano-oscillators[17,24,25] offer high data-throughput and passive synchronization, and can be characterised using simple electrical measurements. The all-electric nature of the input/output to these oscillators has allowed for small artificial neural networks (<10 nodes) to be demonstrated experimentally[26,27], and larger networks have been simulated for RC with binary inputs[28,29]. However, the intrinsic dynamics of single oscillators are relatively simple (though they can be augmented via external delayed feedback[30,31]) and have durations on the order of nanoseconds, limiting their suitability to processing applications where sensory data arrives with characteristic timescales on the order of seconds- far beyond the intrinsic decay times of these systems. Magnetic metamaterials (materials which are engineered to exhibit complex physical responses beyond their underlying material properties) such as artificial spin-ice systems[19,20,32] and skyrmion textures[33], represent an exciting subcategory for magnetic RC, boasting complex, spatially distributed responses. However, interfacing with these materials is challenging, since spin-ices are electrically discontinuous and skyrmion textures require sub-100K temperatures, inhibiting device-tractable measurement approaches.

While there have been many recent, important developments showcasing device-specific RC performance in a range of physical systems, many more general questions remain, such as how different RC architectures can be used to extract different computational properties, and how these architectures can best synergise with the underlying system dynamics. Frequently, the 'single dynamical node' paradigm[34] is employed with little attention to its role in the computation or to the alternative computational properties that could be extracted with different reservoir architectures. This leaves some of the broader potential of nanomagnetic RC as reconfigurable computational platforms untapped.

In this paper, we experimentally demonstrate a pipeline from characterization of device physics, to reservoir design, to state-of-the-art performance in several, diverse computational tasks with a single magnetic device consisting of an array of interconnected magnetic nanorings[35]. The nanoring system boasts a combination of highly complex system response and simple electrical readout: strong coupling between individual ring elements produces complex 'emergent' dynamics (where large-scale responses arise from the collective effects of simple interactions between elements, rather than the properties of the elements themselves), while the continuous nature of the patterned nanostructure facilitates electrical transport measurements. Additionally, non-volatile domain configurations formed in response to input provides a natural means of generating system memory at driving fields an order of magnitude smaller than spin-ice systems[36]. To harness these emergent behaviours, we employ the device in three distinct reservoir architectures that each leverage different aspects of its dynamical properties. We then demonstrate how this provides flexible computational functionality by performing benchmark tasks with contrasting computational requirements on a single device, achieving state-of-the-art accuracies. This highlights the reconfigurability achievable in *in materio* platforms via careful choice of the accompanying RC architecture.

## **Results**

## I- Response of Nanoring Arrays

The devices studied here consist of arrays of 10nm thick $Ni_{80}Fe_{20}$ (Permalloy, Py) nanorings, patterned into a square lattice with each ring having nominal diameters of 4μm and track widths of 400nm, each overlapping with its nearest neighbours across 50% of their track widths[35,36]. The arrays were fabricated by electron beam lithography with lift-off processing and metallised via thermal evaporation. Ti/Au electrical contacts were then added via additional lithography and deposition steps, allowing measurements of the device's anisotropic magnetoresistance (AMR). Despite typical AMR ratios of 3-4% for Py[37], shape anisotropy in the rings means that magnetisation typically runs parallel to the applied currents. This meant only the domain walls which present local changes in magnetisation direction can be detected via AMR, leading to an effective AMR ratio of 0.2% for the device, with the signal quality improved via lock-in amplification techniques (see Methods- Electrical transport measurements). The samples have saturation magnetisation $\mu_0 M_s$ of 0.969 ± 0.006 T, determined via broadband ferromagnetic resonance measurements (see supplementary note 4, and supplementary figure S6). Figure 1a shows a scanning electron microscope image of the device.

In previous studies[35,36], we have shown that interconnected nanoring arrays exhibit emergent magnetization dynamics under rotating in-plane magnetic fields. At the microstate, each ring exists in one of three metastable configurations, defined by the number and position of domain walls (DWs) it possesses, with configurations for 'vortex' (zero DWs), 'onion' (two DWs, 180° separation), and 'three-quarter' (two DWs, 90° separation) shown in figure 1b. To initialise the ring arrays, a strong pulse of magnetic field and subsequent relaxation leads to a uniform state of aligned onion rings, with DWs pointing along the direction of the saturation pulse. Under high driving fields, the DWs can coherently propagate with the applied field, maintaining onion configuration. However, under lower driving fields, stochastic pinning events cause differential movement of DWs within a ring (onion to three-quarter transition), potentially leading to DW annihilation (three-quarter to vortex transition) when itinerant DWs in the same ring collide. DWs can be restored in rings via the propagation of a DW in neighbouring ring, with the magnetic reversal across the junction between the two rings leading to injection of a pair of DWs in the empty ring (vortex to onion/three-quarter transition). Schematics for these processes are shown supplementary figure 2g, 2h. Whilst these behaviours are stochastic at the local scale, interactions between many rings lead to a well-defined global emergent response, providing a complex yet repeatable dynamic state evolution (Figure 1c-1e).

Figure 1: Overview of static and dynamic responses of nanoring arrays

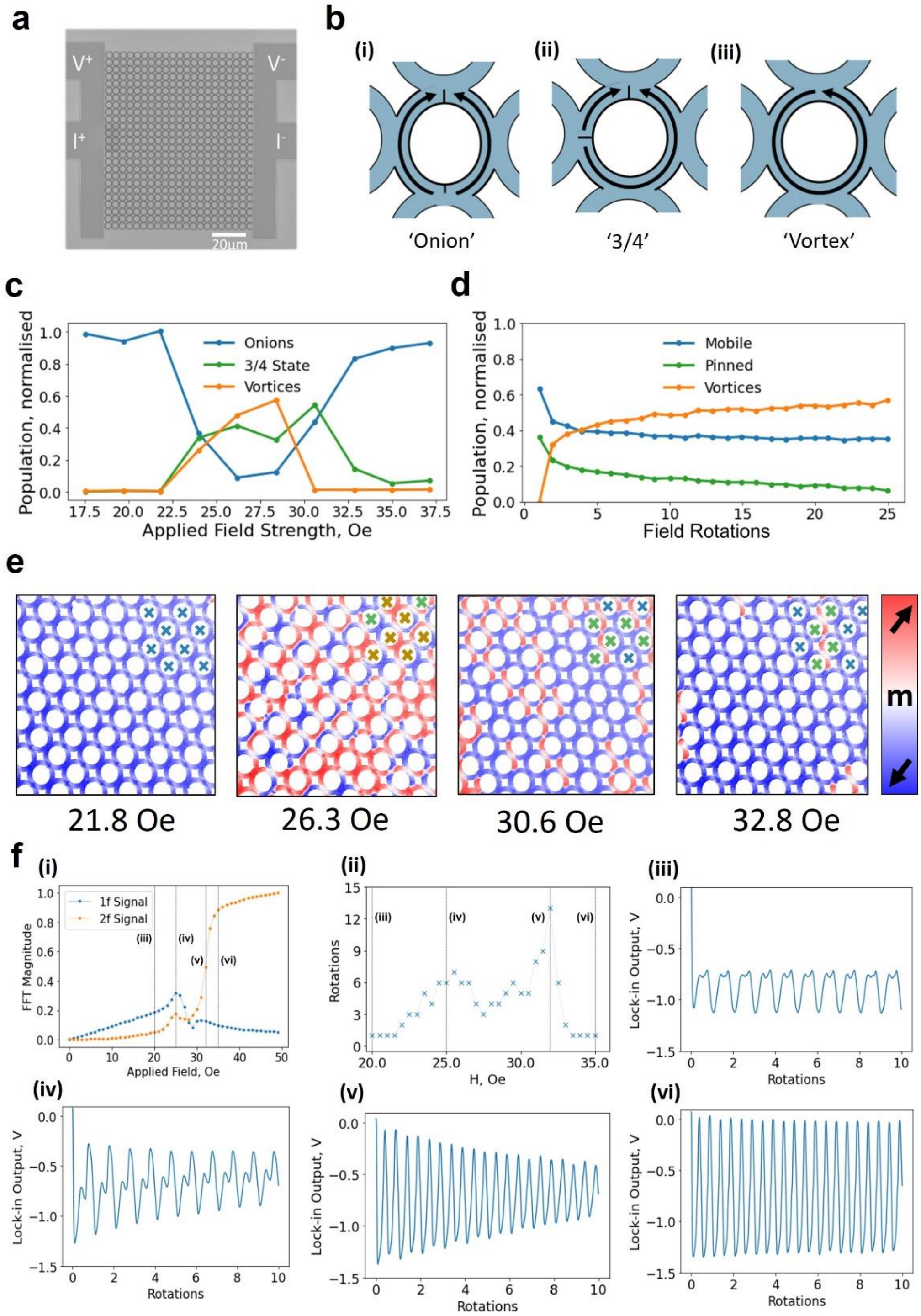

Figure 1a- Scanning electron microscope image showing a nanomagnetic ring array and electrical contacts. 1b- Schematics of available ring domain states, showing (i) Onion, (ii) Three-quarter, and (iii) Vortex. 1c- Varying state population of an array as driving field is increased, taken by counting populations of X-Ray photoemission electron microscopy images (X-PEEM) images after 30 rotations of applied field. 1d- Population of mobile, pinned, and vortex state rings over 25 successive cycles of 27 Oe rotating applied field, inferred from time-varying AMR signals. 1e- X-PEEM images of ring arrays when driven with 30 field rotations of amplitude 21.8, 26.3, 30.6, and 32.8 Oe of applied field. Magnetic contrast is given by the direction arrows on the colour bar, crosses in the top right corner rings denote (blue) onion, (orange) vortex, and (green) three-quarter ring configurations. 1f(i)- Fourier components of AMR signal of arrays driven with 10 rotations of magnetic field at various applied fields. Blue datapoints show Fourier component at the driving field frequency (1f), while orange datapoints show component at double the driving field frequency (2f). (ii)- Number of rotations of field required for the system to reach an equilibrium response (<2% peak-to-peak deviation between cycles) from saturation for a range of applied fields. (iii)-(vi) - Measured lock-in voltage of the array when driven with 10 field rotations of amplitude (iii) 20 Oe, (iv) 25 Oe, (v) 32 Oe, and (vi) 35 Oe from saturation.

To evaluate the evolving magnetic states of the arrays for computation, AMR measurements performed via the electrical contacts shown in Figure 1a. This gives a single global readout for each array, which varies over a given input rotation. Initially, the device's response as a function of rotating field amplitude was surveyed to determine the characteristics of the responses and identify computationally useful features (Figure 1f). Fourier analysis of the AMR response led to observation of two distinct signals with frequencies that match (1f signal) as well as double (2f signal) the frequency of the rotating magnetic field, with the relative magnitude of the two signals with respect to driving field amplitude shown in Figure 1f(i) (see Supplementary note 1 for further Fourier analysis). Physically, these processes can be separated into elastic deformation of the rings' domain structures due to susceptibility effects (1f, dominant at lower fields), and irreversible DW propagation between pinning sites in the rings (2f, dominant at higher fields). Further details of these mechanisms can be found in the supplementary note 2. The dynamic nature of the system's response was evaluated by measuring the number of rotations that were required for the AMR signal to reach dynamic equilibrium (<2% amplitude variance between cycles) from saturation, with the measured timescales and the underlying signals shown in figures 1f(ii), and 1f(iii) – 1f(vi) respectively. The onset of DW motion can also be observed at a ~22 Oe, marked by the nonlinear increase of 1f signal in figure 1F(i), as well as the start of varying time-signals between cycles in figure 1f(iii) – 1f(vi).

From these measurements, three computationally promising properties can be identified. Firstly, the distinct variation of the AMR frequency components with respect to field provides crucial nonlinearity. Secondly, the dependence of the device's response on its past states, as evidenced by the range of timescales observed in the AMR signals, allows information to be connected across time in manner reminiscent of the echo-state property of echo state networks (ESNs). Finally, the presence of a threshold field below which no irreversible DW motion occurs shows a non-volatility of system state, providing pathways to longer-term storage of information.

The key demonstration of this paper is how these physical behaviours of the nanoring devices can be harnessed in different ways to create RCs with different computational properties, and thus tackle problems with different computational requirements. We achieve this by incorporating the device into three distinct reservoir architectures: an approach which takes advantage of the time-continuous oscillations of the nanoring array (signal sub-sample reservoir), the 'single dynamical node' architecture introduced by Appeltant et al.[34], and the recently proposed 'rotating neurons reservoir' of Liang et al.[38], yet to be deployed outside of analogue electronic RC. These architectures are presented schematically in Figure 2 and described in their respective Methods sections. In the following, we will explain how each of these architectures allows different computational properties to be emphasised and then exploited to perform challenging computational tasks. For further details on the methods employed for the machine learning tasks, see Supplementary note 5.

## II- Signal Sub-sample Reservoir

Figure 2- Schematic diagrams of each reservoir architecture

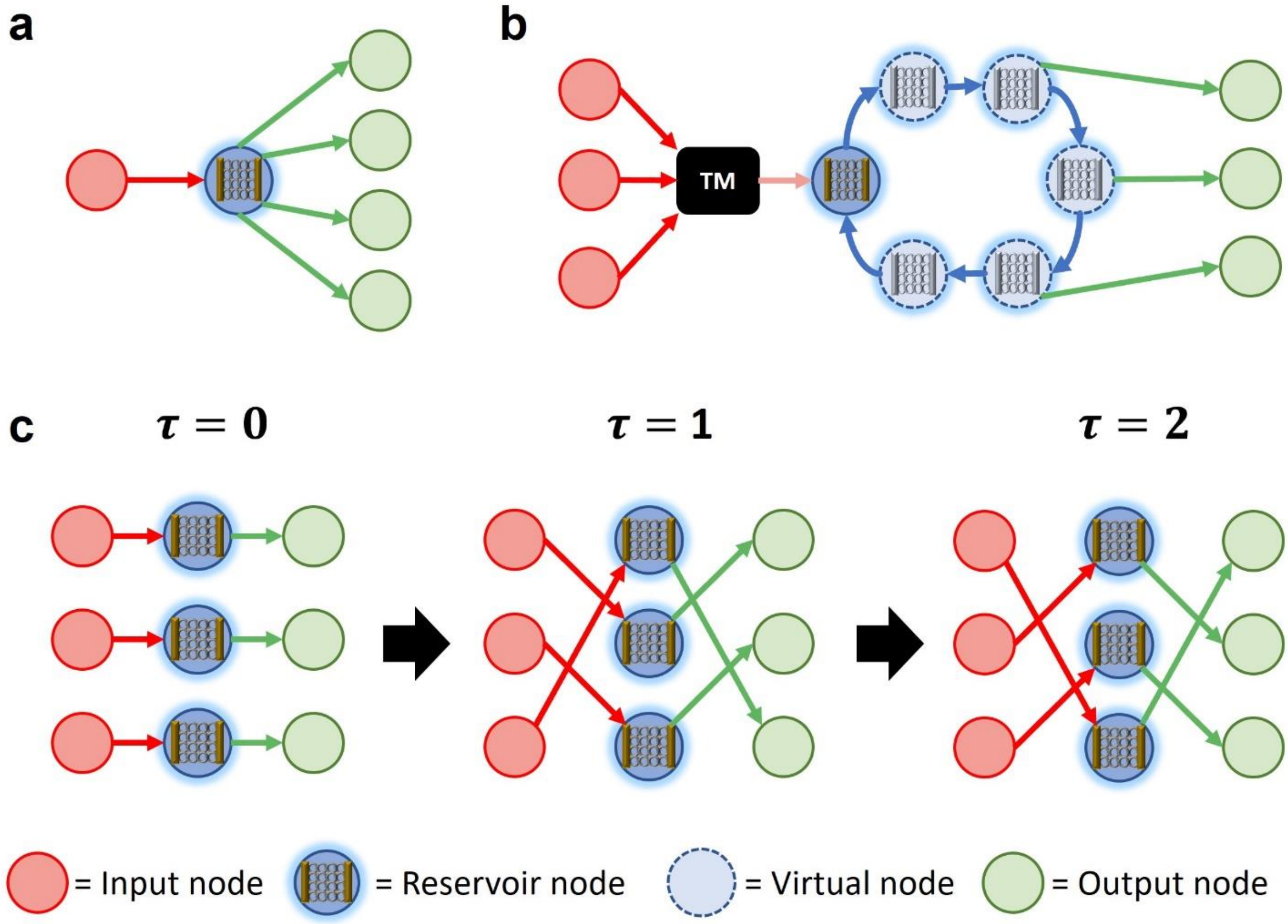

Figure 2- Schematic diagram showing three different reservoir architectures (a-c), with differing methods for providing input data (red circles) into reservoir nodes (blue circles) and reading reservoir state as output (green circles). a- Signal sub-sample architecture, showing a singular input datum fed into the ring arrays, with multiple state readouts taken from the single node. b- Single dynamical node architecture, where multiple input dimensions are time-multiplexed (black rectangle), before being fed into a single node. 'Virtual nodes' (pale blue circles), are generated from the dynamical node as input varies over time, generating outputs for each virtual node. c- Revolving neurons architecture, where the weighted connections between input-to-reservoir and reservoir-to-output change consecutively with each input timestep $\tau$.

One foundational task for RC platforms is nonlinear signal transformation[20,32,39–41]. In this problem, the system is provided with input of a given periodic response and is tasked with transforming the input signal into a different target signal. To perform this task, the reservoir should provide a higher-dimensional, nonlinear representation of the input signal so that the transformation between the input and the target can be computed via a simple linear readout.

To meet these computational demands, we designed a simple reservoir input/output architecture that directly exploited the non-linear variation of the 1f/2f frequency signals (Figure 3a). Here, each input datum scaled the field amplitude for a single rotation, and the resulting AMR response was sampled at 32 times per input, expanding input dimensionality 32-fold. The two frequency components have different nonlinear variations with respect to input magnitude, meaning that the relative magnitude of the continuous signal at fixed sample points will have nonlinear variation with respect to each other, providing dimensionality expansion of the input data. This offers a very simple method for providing increased nonlinearity in physical systems with continuous signals, obtained by leveraging a phase transition in system response.

Figure 3b-3d shows the resulting signal reconstruction when the ring array system was tasked with transforming sinusoidal input to ReLU(sin(x)) (rectified linear unit), square wave, sawtooth waveforms. To evidence the impact of the metamaterial on computation, a control experiment was performed by recording the voltage of one of the driving electromagnets as the measured reservoir state instead of the resistance of the nanoring array. This provided equal dimensionality expansion as the nanoring array transformation, but without the nonlinearities contributed by the nanoring system. However, these measurements do contain any hardware-based nonlinearities in the electromagnets such as slew-rate between inputs and inductive effects, accounting for any nonlinearities provided by the experimental equipment. The ring array network outperformed the control network in all cases, offering up to a 55-fold reduction in MSE ($4.6\text{x}10^{-4}$ compared to $2.5\text{x}10^{-2}$) when replicating the ReLU function. The rings also perform favourably compared to proposed spin-ice platforms, with lower errors for Sawtooth ($1.406\ \text{x}10^{-2}$ vs $1.919\text{x}10^{-2}$) and Square ($6.605\ \text{x}10^{-3}$ vs $2.429\ \text{x}10^{-2}$) waves[20]. The different reconstruction tasks are performed optimally at very different ranges of applied field (Figure 3e), highlighting how the ring array's dynamics can further tuned to for better performance in a range of similar problems even when held within a consistent reservoir architecture. The accuracies for all transformations for both the ring array and control network, as well as the ratio between them, are shown in the figure 3f.

Figure 3- Performance of signal transformation task

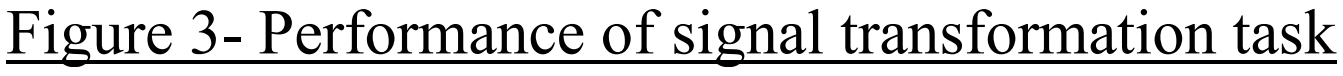

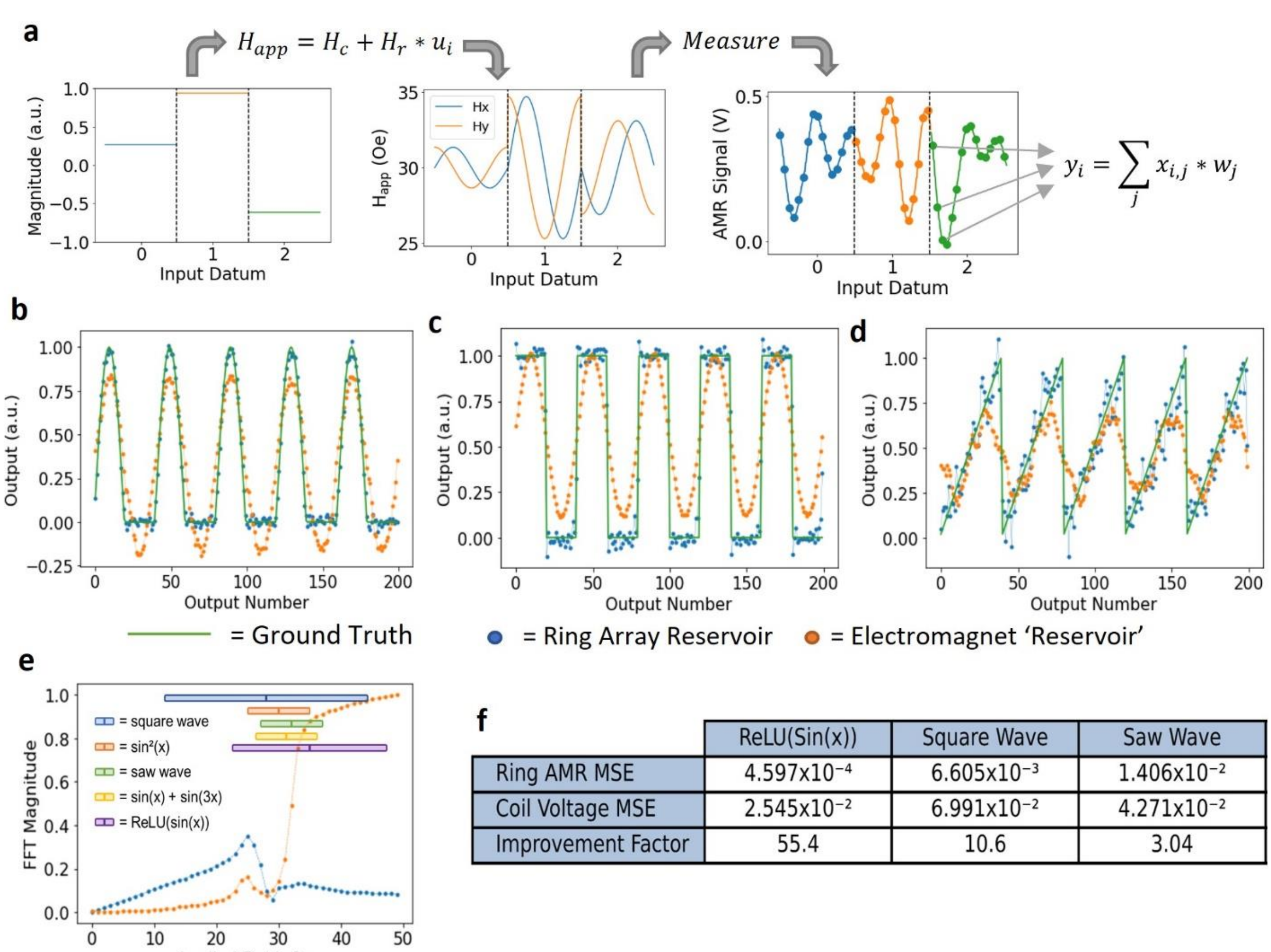

| | ReLU(Sin(x)) | Square Wave | Saw Wave |
|---|---|---|---|
| Ring AMR MSE | $4.597\text{x}10^{-4}$ | $6.605\text{x}10^{-3}$ | $1.406\text{x}10^{-2}$ |
| Coil Voltage MSE | $2.545\text{x}10^{-2}$ | $6.991\text{x}10^{-2}$ | $4.271\text{x}10^{-2}$ |
| Improvement Factor | 55.4 | 10.6 | 3.04 |

Figure 3a- Schematic diagram showing scaling of input data $u_i$ to applied field $H_{app}$, application of field rotations as components of field in x and y, $H_x$ and $H_y$ respectively, followed by sampling of resulting anisotropic magnetoresistance (AMR) signals to produce features, combined via a weighted sum to produce output. 3b-f: optimal reconstructions obtained from the Ring Array (blue) as well as the control measurements of electromagnet voltage (orange) compared to the desired target signal (green), for (b) rectified linear unit(sin(x)), (c) square wave, (d) saw wave. 3e- Input scaling parameters corresponding to reservoir configurations with minimum error for the signal reconstruction task, overlayed on relative 1f (blue dotted line) and 2f (orange dotted line) signal magnitudes over a range of applied fields. Bar width demonstrates

applied field range ($H_r$), with central field ($H_c$) marked by the solid line. 3f- Comparison of mean-squared error between target signal and reconstructions drawn from the measured Ring Array voltages, as well as a control measurement taken from voltage measurements of the driving electromagnets.

## III- Single Dynamical Node Reservoir

Another key application for RC is the classification of time varying signals such as spoken digits, a task which has been previously used to benchmark a variety of RC platforms[17,21,25,34]. While input data for the previous task was 1-dimensional, input data for speech recognition tasks are typically multi-dimensional. Furthermore, non-linear interactions between these input dimensions in the reservoir are essential to successful classification. Here, we consider classification of the spoken digits 0-9 from the TI-46 database (see Supplementary Note 5- Spoken Digit Recognition Task for details). The input data was 13-dimensional, consisting of the results of applying Mel-frequency cepstral filters[42] to each utterance. The data is linearly inseparable, with classification accuracy being limited to around 75%[25] if input data is passed directly to a linear readout layer. The role of the reservoir is to provide a non-linear mapping of input data into higher dimensional reservoir space, thus allowing the linear readout layer to establish hyperplanes which can classify the data accurately.

Tackling this problem requires a reservoir architecture that expresses the non-linearity of the device's AMR response, can accommodate multiple input dimensions, and allows nonlinear combinations of these input dimensions, properties that cannot be provided by the signal sub-sample architecture. To satisfy these requirements we adopted the single dynamical node approach (Figure 4a) initially proposed by Appeltant *et. al.*[34] and detailed in the Methods section. Multidimensional input data was fed sequentially into the device, creating a reservoir constructed of 'virtual' nodes that convolves inputs temporally via the ring array's transient dynamics. Thus, this approach leveraged both the non-linear response of the device's AMR signal to input (via the activation of the virtual nodes), and its transient nature (which allowed interaction between virtual nodes).

As shown in the previous task application, our device exhibited a broad range of responses that were potentially useful for computation. Searches over parameter space can be performed for simple tasks such as signal transformation, however for more data-intensive tasks, this process is inefficient. Previous studies have shown that task-agnostic metrics, which can be found via statistical analysis of small random datasets[36,43], can speed up parameter selection by identifying promising regions of parameters space. Using metrics of kernel rank (KR, the ability of the reservoir to separate different input classes) and generalization rank (GR, and the ability of the reservoir to generalise inputs of the same class), we evaluated the computational properties of the device's transformations for a range of scalar parameters controlling the scaling ($H_r$) and offset ($H_c$) of inputs (see supplementary note 3). As the spoken digit recognition task required improving the linear separability of input data, KR was chosen to be the key identifier of promising performance, with a comparatively lower GR also needed to generalise between the different speakers.

To highlight the single dynamical node approach's better suitability to the spoken digit recognition task, metric maps were also drawn similarly for the other reservoir architectures (Supplementary figure S4). While the revolving neurons reservoir showed good separation properties (high KR), and the signal sub-sample reservoir good generalisation properties (low GR), only the single dynamical node architecture exhibited a balance of the two, showing better suitability for classification tasks. This is likely due to the rotating neuron reservoir's increased dependency upon past states reducing its ability to generalise, and the relatively smaller dimensionality of the signal sub-sampling reservoir leading to poorer ability to separate information. Conversely, the single dynamical node approach both provides good dimensionality expansion, as well as having decreased dependence on past states via the increased separation of inputs over time provided by the time multiplexing procedure.

Figure 4b shows the error rates versus training samples for spoken digit recognition, obtained using both a ‘promising’ ($H_c$ = 29 Oe, $H_r$ = 10 Oe, KR = 76) and arbitrarily chosen reservoir configurations (e.g., $H_c$ = 21 Oe, Scaling, $H_r$ = 7.5 Oe, KR = 52). 100-fold cross-validation was performed to evaluate general performance and find a suitable regularization parameter, selected for best performance on the training set to prevent overfitting. Again, a reservoir constructed from the voltage signals across the driving coils was used as a control, effectively skipping the reservoir transformation whilst including the same pre-processing steps. A significant reduction of word-error-rate is observed moving from ‘control’ to ‘arbitrary’ to the ‘promising’ case, with error-rates of 24.8%, 10.4%, and 4.6% respectively. This demonstrates not only the effectiveness of the reservoir’s transformations in improving the linear separability of the data, but also the utility of evaluating metric scores to expedite system parameter selection.

One method for further improving performance commonly employed in conventional RC settings is the use of bespoke learning rules instead of standard regression-based training methods. Here, the SpaRCe[44] algorithm was used, which was developed for use on ESNs though thus far has not been applied to physical systems, and its online nature synergises well with life-long learning paradigms especially useful for system-level device applications[45]. The algorithm aims to suppress confounding information and induce sparse output representations. Here, these properties help to mitigate the effects of experimental noise and remove redundant virtual node outputs. With SpaRCe, the accuracy was improved to 99.8%, as shown in Figure 4c. The ring arrays matched state-of-the-art performance compared to other magnetic architectures, even with fewer (50) virtual nodes used in the time-multiplexing procedure (STNOs with 400 virtual nodes, 99.8%[17], simulations of superparamagnetic arrays with 50 virtual nodes, 95.7%[21]), and improved upon the performance achieved in simulations of the ring system (97.7%[36]).

Figure 4- Performance of spoken digit recognition task

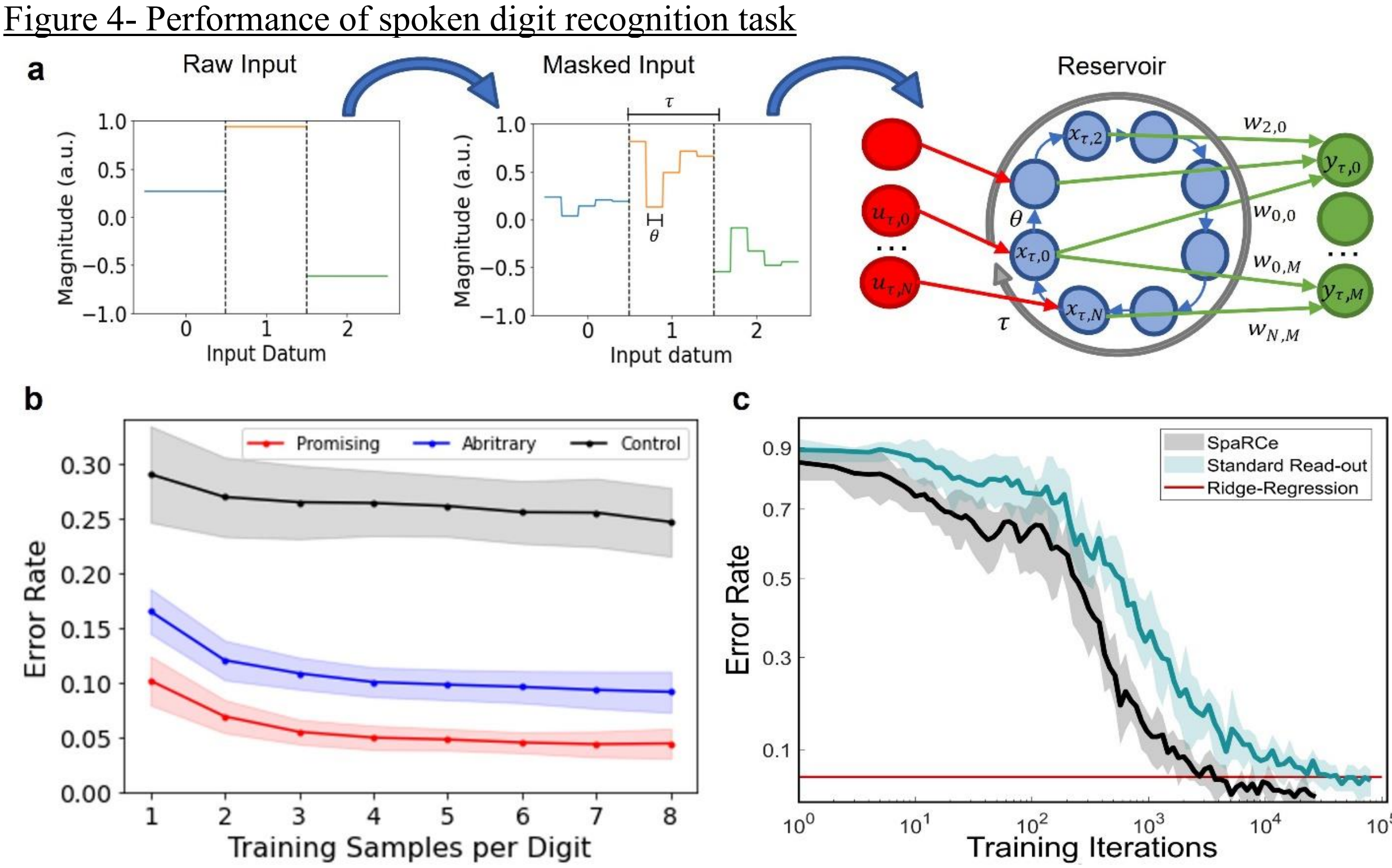

Figure 4a- Process showing time multiplexing procedure, taking raw inputs, combining them with a fixed mask to produce masked input (each of N virtual nodes has an input of duration θ, totalling to a duration of $\tau = N\theta$ per unmasked input), then inputting those inputs to the reservoir and measuring evolving reservoir state. 4b- Error rate versus number of sequences used for training for ‘promising’ ($H_c$ / $H_r$ = 29±10 Oe, (red)) and ‘arbitrary’ ($H_c$ / $H_r$ = 21±7.5 Oe, (blue)) reservoir parameters, and control measurements taken from voltage

readings of the input electromagnets for the 'promising' case (black). The shaded regions show the standard deviation of performance over the 100-fold cross-validation. 4c- Error rate vs training iteration comparison between online learning methods using the SpaRCe algorithm (black) and standard online learning (blue) for a system driven with $H_c$ and $H_r$ values of 29±10 Oe. The shaded region shows minimum and maximum accuracies over 10-fold cross-validation. Red line shows accuracy achieved with ridge regression.

## IV- Revolving Neurons Reservoir

In addition to data classification tasks, RC is also highly applicable to time series prediction problems. To be successful in these tasks, RC platforms often require fading memory of past inputs to correctly predict future trajectories, in addition to the non-linear properties that were exploited in the previous two tasks. The memory of a reservoir can characterised by evaluating the linear memory capacity[46] (MC), which measures the ability to reconstruct past inputs from the current reservoir state over increasing delays. Typically, nanomagnetic RC platforms exhibit low MC without the inclusion of delayed feedback due to the short timescale of intrinsic dynamic behaviors[23]. Additionally, reservoirs constructed under the single dynamical node paradigm struggle to recall previous input datapoints due to the long temporal separations between each input created by the time-multiplexing procedure. For example, the prior architectures presented here exhibited peak MC < 3, meaning they could only reliably recall the previous two inputs (see supplementary figure S5).

To utilise the system's non-volatile properties and create a architecture better suited to time series prediction tasks, the recently proposed 'rotating neurons reservoir'[38] (RNR, see Methods) configuration was employed. Here, the system was constructed from 50 distinct dynamical nodes, with inputs to each node modulated by a fixed, rotating input (output) mask which multiplied input (reservoir state) values by ±1 (weight value), shifting the input/output connections to each node by one position every timestep (Figures 2c, 5a).

The memory effects exhibited in this configuration emerge from the ratchet-like nature of the device's non-volatile response: small inputs cause reversable perturbations while larger inputs cause non-volatile changes to underlying domain structure. In the RNR configuration, this means the system's evolution is dependent upon the sign of the input at a given time, determined by the mask. For negative mask values, the low applied field strengths leave the rings' domain structures unchanged through multiple timesteps until a positive input is applied to the system, where the higher applied fields cause DW propagation which is then measured as a change in the system's resistance. This allows inference of the previous inputs applied to the system from the current states of the dynamical nodes, increasing MC. This architecture hence synergises well systems where activity decays slowly in the absence of large inputs.

Figure 5b shows the MCs calculated from the ring array system using the RNR approach. A peak MC of around 11.5 was found at $H_c$ = 21 Oe and $H_r$ = 10 Oe, showing that the device's non-volatile properties were being harnessed to provide much greater memory of past inputs than the other approaches, and thus extending applicability to problems with longer-term temporal dependencies. The region of maximum MC here is correlated to the central field at which DW motion starts to occur (Figure 1f(ii)). This corroborates the reasoning that the movement of DWs into different non-volatile configurations at fields above this value is where the system is 'storing' its memory of past states.

While MC can quantify the extent of linear memory (direct reconstruction of past inputs) in the system, real-world regression problems often require nonlinear memory (nonlinear representations of past inputs) for accurate prediction. To demonstrate the extent of nonlinear memory available to the system, we trained the system to reproduce a nonlinear auto-regressive moving average (NARMA-N) of input signals with varying degrees of autocorrelation (NARMA-5 and NARMA-10). For this problem, a system with perfect linear memory of equal degree to the autocorrelation (i.e., a shift-register of length N) can only achieve normalised means squared errors (NMSE) of around 0.4[34]. To improve upon this, a system needs to store nonlinear representations of past inputs. Figure 5c, 5d presents heatmaps of NMSE achieved over a range of field scaling parameters for NARMA-5 (5c), and NARMA-10 (5d), as well as examples of the reconstructed signals (5e, 5f). Regions where the ring array system outperforms the shift register in the NARMA-5 and NARMA-10 tasks are shown by the grey lines in Figure 5c, 5d, achieving peak NMSEs of 0.265 and 0.359 respectively. The combination of MC and performance of NARMA-N demonstrated that the system had been effectively reconfigured into a configuration with both linear and nonlinear memory without the aid of external delayed feedback lines that have typically been used in other demonstrations.

Figure 5- Performance of linear and nonlinear memory tasks.
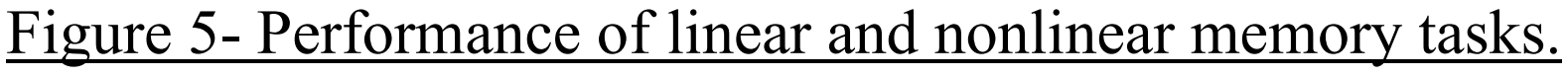

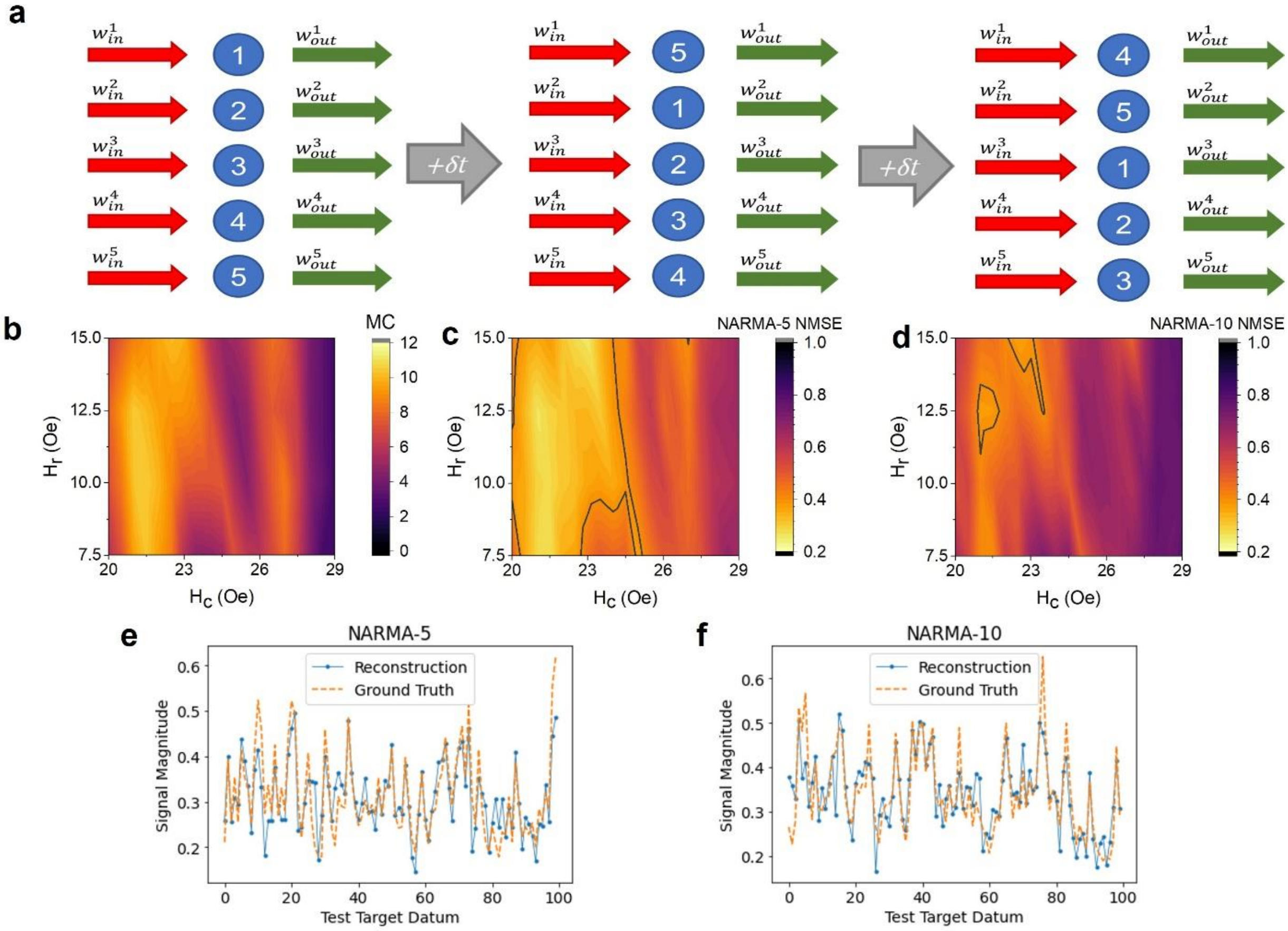

Figure 5a- Schematic diagram for simplified revolver setup consisting of three nodes, showing how input (red arrows) and output (green arrows) change with each timestep τ with respect to fixed dynamical nodes (blue circles). 5b- Memory Capacity (MC) over a range of field scaling parameters under the rotating neurons reservoir construction. 5c/d - Performance heatmaps for the (c) nonlinear autoregressive moving average (NARMA) -5 and (d) NARMA-10 system approximation task. Regions inside the grey line show configurations outperforming the score of a shift register with equal degree to the NARMA problem. 5e/f – NARMA signal reconstruction for optimally performing ring array reservoirs (blue, Normalised mean squared error (NMSE) = 0.265) compared to ground truth (orange, NMSE = 0.359) for (e) NARMA-5 and (f) NARMA-10.

## Conclusion

In this paper we have demonstrated how a range of different RC architectures allow exploitation of different underlying dynamic properties in a complex magnetic system. This reconfigurability allowed the platform to achieve state-of-the art performance in three diverse tasks with differing computational requirements. To summarise the key correlations between underlying dynamics and suitable reservoir architectures, we found that the signal subsampling architecture synergises with phase transitions in the system's response to provide nonlinear mappings of input, the single dynamical node paradigm synergises with transient responses to connect different input dimensions across time, and the rotating neurons reservoir scheme synergises well with regimes where reservoir state changes slowly with small/zero inputs, allowing information from past inputs to be sustained over time via the rotating input mask.

The synergy between these dynamic properties is also directly correlated to the type of task that the resulting reservoir is suitable for solving: the dimensionality expansion and nonlinear dynamics provided by the signal subsampling architecture allows for effective 1D signal processing, the temporal mixing of input dimensions in the single dynamical node architecture enables classification on multivariate data, and the slow dynamics modulated by the rotating input mask in the rotating neurons reservoir architecture allows for effective performance in memory based tasks. Aside from the architecture choice, the selection of suitable scaling parameters for the input data is also critical to performance. To address this, we used task-independent metrics to provide a more holistic mapping of the computational properties of the reservoir across a range of scaling parameters and demonstrated the additional performance attainable via selecting promising parameters from the resulting metric maps for both classification-based tasks (KR/GR) and memory-based tasks (MC), with additional comparisons between each of the architectures' scores in these metrics.

We believe that the range of dynamical regimes offered by the system, combined with the ability to address each of these properties separately and extract distinct computational properties via controlling the external reservoir architecture, makes the ring system a candidate for reservoir computing with complex dynamic substrates. Additionally, the effectiveness of synergising the reservoir architecture with the dynamic properties of the underlying system makes for an effective methodology for extracting a broad range of computational capability for other similar devices. The ring devices are not without their limitations however, with the current device being driven external rotating magnetic fields, which provides both a limitation on the throughput on data input to the system (on the order of 100s of Hz), and power wastage in generating the magnetic fields over areas orders of magnitude larger than the nanoring array itself. Additionally, the current electrical readout provides a single scalar readout on the entire system state at a given point in time, which is sub-optimal for extracting complex state information on a system which exhibits spatially distributed responses like the ring system here. The feasibility of the ring system as a complex RC device that would be applicable to real-world settings hinges upon the ability to respond to electrical inputs such as spin-orbit torque driven DW motion, as well as expanding upon the readout mechanism to provide spatially resolved measurement of magnetic state.

To expand the computational capabilities of the ring arrays, the complex behaviours outlined here should operate concurrently as part of a larger system. The changes in magnetic responses offered via geometric changes to the system could enable multiple devices to operate in different regimes of dynamics and emphasise different computational properties under a single input field. Other magnetic metamaterial platforms have been shown to be useful in 'deep' reservoir networks with distributed reservoir properties[32], which the ring system would also likely benefit from. We believe that this work marks a significant step forward towards the realisation of metamaterial systems as computational platforms that are device-compatible, and that the rich playground of computationally useful dynamics they offer makes the ring system a promising candidate for physics-based neuromorphic computation platforms.

## Methods

### Device Fabrication

The ring array devices were fabricated using two-stage electron-beam lithography, with layouts patterned using a RAITH Voyager system. Wafers of Si (001) with a thermally oxidized surface were spin-coated with a positive resist. The ring structures were metallized to thicknesses of 10nm via thermal evaporation of $Ni_{80}Fe_{20}$ powder using a custom-built (Wordentec Ltd) evaporator with typical base pressures of below $10^{-7}$ mBar. The initial resist went through lift-off, leaving the ring structures before re-application of the resist and further electron-beam lithography. Electrical contacts were metallized in two stages of thermal evaporation, first with 20nm titanium to form a seed layer, before growth of 200nm of gold. Electrical connections were provided between the device and a chip carrier through bonding of gold wire between contact pads on the device and the chip carrier.

Electrical Transport Measurements of Ring Arrays

Currents of 1.4mA were provided to the arrays as a 43117 Hz sine wave into the patterned contacts (Figure 1a) on the device using a Keithley 6221 current source. Resistance changes via AMR effects were measured using a Stanford Research SR830 lock-in amplifier. A National Instruments NI DAQ card was used to log the output voltage of the lock-in amplifier 64 times per rotation of applied field, and the data were then saved on a personal computer. The rotating magnetic fields were generated using two pairs of custom-built air-coil electromagnets in a pseudo-Helmholtz arrangement. The electromagnets were driven by a pair of Kepco BOP 36-6D power supplies and were controlled via voltage signals generated using a computer and the analogue output functionality of the NI card. A rotating field frequency of 37 Hz was chosen as a compromise between data throughput and signal fidelity.

Reservoir Computing

In RC, the fixed reservoir layer provides a transformation of discrete-time input signals $u(t)$, to reservoir states, $x(t)$, according to the internal dynamics of the reservoir layer. The readout layer (here, a single-layer linear perceptron) provides a weighted sum of the reservoir states as output, $y(t)$. The transformation provided by the reservoir layer results in a higher-dimensional mapping of the input signals. This aids the output layer in classifying the input signals by allowing selection of hyperplanes in higher-dimensional space to correctly classify data that was previously linearly inseparable.

In this work, the RNN that constitutes the reservoir layer of the typical echo state network (ESN) was replaced with the magnetic nanoring device. The reservoir transformation was provided by the physical processes that govern the array's magnetic response to field, as well as the changes to electrical resistance that consequently occur. Methods for inputting and extracting data are outlined for each reservoir configuration:

Signal Subsample Reservoir

Input sequences $u_\tau$ are transformed to give an applied field sequence via a pair of scalar parameters $H_c$ and $H_r$, shown in the following equation, which represent the zero-input field offset and the field scaling factor respectively:

$$H_{input} = H_c + H_r * u_\tau$$

Each input was applied for a single rotation of magnetic field. The reservoir states were then extracted by sampling the lock-in voltage signal 32 times per rotation, producing a 32-node output.

Single Dynamical Node Reservoir

This approach uses 'virtual' nodes[34], where the reservoir states are generated from observing the state of the nanoring array as it evolves under time-multiplexed input. The generation of the time-multiplexed sequence of applied field magnitudes, $H_{input}$, (a vector of length $\theta * \tau$, where θ represents the desired number of virtual nodes, and τ the number of discrete-time windows the initial input sequence contains) was accomplished by combining the d-dimensional input vector for each timestep in $u_{\tau d}$ with a fixed input mask matrix, $M_{d,\theta}$, and flattened into a 1D sequence by concatenating timestep-by-timestep via:

$$H_{input} = H_c + H_r \sum_{k=1}^{\tau} u_{k,d} * M_{d,\theta}$$

where $M_{d,\theta}$ consisted of randomly generated 0's and 1's. The field sequence was then input to the system by rotating the field at magnitudes specified by $H_{input}$ for a given number of quarter-rotations per input datum.

The resulting voltage signals provided by the lock-in amplifier underwent some simple processing steps: Firstly, a high-pass filter with a low cut-off frequency of 3Hz was used to centre the signals about zero and remove any low-frequency noise in the system. Band-pass filters were used to extract the 1f and 2f components separately. The pass-windows for each of these filters were centred about the input frequency and twice the input frequency, with band widths of 25% of the centre frequency to capture the damped dynamics of the oscillatory system. The outputs of the high-pass, and each of the band-pass filters, were sampled twice per input, forming a complete reservoir state vector six times the length of $H_{input}$.

### Rotating Neurons Reservoir

This technique employs a shifting input/output mask[38], functionally analogous to rotating the input and output weights synchronously while keeping the dynamical neurons fixed. The procedure for this 'rotation' can be described as follows: Consider a system of $\theta$ dynamical nodes $\eta^i$, where i denotes the index of each node. An input signal $u_{\tau,d}$ is combined with mask $M_{d,\theta}$, to produce input dimensions $s_{\tau\theta}$. The input to node $\eta^i$ at timestep t, $\tilde{s}_{t,i}$, is given via

$$\tilde{s}_{t,i} = s_{t,(i+t)\%\theta}$$

where '%' represents the modulo operation. The resulting output matrix, $\tilde{X}_{\tau\theta}$, is generated by vertically concatenating the output of vectors all $\theta$ nodes as they evolve, and is 'unraveled' similarly to form reservoir state matrix $X$ via:

$$X_{t,i} = \tilde{X}_{t,(i-t)\%\theta}$$

Additional information on each of the machine learning tasks, details of training methods employed, and any data processing steps taken can be found in Supplementary Methods - Machine Learning Tasks.

### Data Availability

A repository containing all data used in this study can be found on the University of Sheffield's online database, ORDA, DOI: 10.15131/shef.data.23815434

### Code Availability

A repository containing all code used in this study can be found on the University of Sheffield's online database, ORDA, DOI: 10.15131/shef.data.23815434

### Author Contributions

ITV performed all experiments, analysis, and modelling, and drafted the article. ITV and CS designed and optimised the AMR measurement apparatus. ITV implemented and performed all machine learning frameworks excluding the SpaRCe algorithm, implemented by LM. GV, CS, and PWF designed and manufactured all devices. ITV, AW, RMRR, DAA, and TJH performed X-PEEM measurements, which were overseen by DB, FM, and SSD. EV, DAA, and TJH conceptualised and supervised the work. All authors reviewed the manuscript.

Acknowledgements:
The authors would like to thank Jack C. Gartside for helpful discussions. The authors thank STFC for beam time on beamline I06 at the Diamond Light Source, and thank Jordi Prat, Michael Foerster and Lucia Aballe from ALBA for providing quadrupole sample holders[47]. I.T.V. acknowledges a DTA-funded PhD studentship from EPSRC. The authors gratefully acknowledge the support of EPSRC through grants EP/S009647/1, EP/V006339/1, and EP/V006029/1. This project has received funding from the European Union's Horizon 2020 FETOpen program under grant agreement No 861618 (SpinEngine). This work was supported by the Leverhulme Trust (RPG-2018-324, RPG-2019-097).

Competing Interests
The authors have no competing interests to declare.